\documentclass[aps,prd,twocolumn,showpacs,amsmath]{revtex4}
\usepackage[dvips]{color,graphicx}
\usepackage{amsfonts}
\usepackage{amssymb}
\usepackage{latexsym}

\newcommand{\dalm}{\kern1pt\vbox{\hrule height 0.9pt\hbox{\vrule width
0.9pt\hskip 2.5pt\vbox{\vskip 5.5pt}\hskip 3pt\vrule width 0.3pt}\hrule height
0.3pt}\kern1pt}
\newcommand{\ma}[1]{\mbox{$\mathcal{#1}$}}

\newcommand{\lw}[1]{\smash{\lower2.ex\hbox{#1}}}

\begin{document}

%\thispagestyle{empty}

%<<<<<<<<<<<<< TITLE >>>>>>>>>>>>>>>%
\title{On Kaluza-Klein spacetime in Einstein-Gauss-Bonnet gravity }
%<<<<<<<<<<<<< AUTHOR >>>>>>>>>>>>>>>%
\author{Alfred Molina$^{1}$\thanks{Electronic address:alfred.molina@ub.edu}
and
Naresh Dadhich$^{2}$\thanks{Electronic address:nkd@iucaa.ernet.in}}

%<<<<<<<<<<<<< ADDRESS >>>>>>>>>>>>>>>%
\affiliation{
$^{1}$Departament de Fisica Fonamental, Universitat de Barcelona, Spain\\
$^{2}$Inter-University Centre for Astronomy \& Astrophysics, Post Bag 4, Pune~411~007, India\\
}
\date{\today}

%======================================%
%<<<<<<<<<<<<< ABSTRACT >>>>>>>>>>>>>>>%
%======================================%
%======================================%
%\begin{abstract}
\begin{abstract}
By considering the product of the usual four dimensional spacetime with two 
dimensional space of constant curvature, an interesting black hole solution has recently been found for Einstein-Gauss-Bonnet gravity. It turns out that this as well as all others could easily be made to radiate Vaidya null dust. However there exists no Kerr analogue in this setting. To get the physical feel of the four dimensional black hole spacetimes, we study  asymptotic behavior of stresses at the two ends, $r\to0$ and $r\to\infty$.
\end{abstract}

%<<<<<<<<<<<<< PACS NUMBER >>>>>>>>>>>>>>>%

\maketitle

\section{Introduction}
We consider the spacetime with the topology of ${\ma M}^4 \times {\ma K}^{n-4}$, where ${\ma K}^{n-4}$ is the $(n-4)$-dimensional space of constant curvature and ${\ma M}^4$ is the usual four dimensional spacetime. It turns out that the vacuum equations with Gauss-Bonnet contribution and $\Lambda$ also split up into $4$ dimensional part and $(n-4)$ extra dimensional part. The former is equivalent to the usual Einstein equation with redefined $\Lambda$ which will have the general solution as Schwarzschild-AdS/dS. When it is put into the latter scalar constraint given by the extra dimensional equation, Schwarzschild mass has to vanish and the solution then reduces to AdS/dS. However, there is an exceptional case where $4$-dimensional equation is made completely vacuous by proper prescription of constant curvature of $(n-4)$-space and $\Lambda$ in terms of Gauss-Bonnet parameter $\alpha$. That is $4$-metric remains completely free and it is entirely determined by the scalar constraint which admits the general solution giving the new black hole ~\cite{md1}. It is an interesting solution which asymptotically goes over to Reissner-Nortstr\"om-anti-de~Sitter (RN-AdS) spacetime in spite of absence of Maxwell field. It is a black hole which is purely created by coupling of extra dimensional curvature and $\Lambda$ with $\alpha$. The prescription requires Kaluza-Klein split up of spacetime and both Gauss-Bonnet term as well as $\Lambda$. In this case, the base spacetime was spherically symmetric and new solution is Schwarzschild analogue. There is a general feature of this setting that all the static solutions in four and lower dimension could always be generalized to Vaidya radiation. \\

There exists extensive literature on vacuum solutions of Einstein-Gauss-Bonnet gravity in higher dimension as well as in the brane-bulk system. The fine-tuning of $\Lambda$ and $\alpha$ was first investigated in ~\cite{whe} and it turned out that in some cases the metric contained an arbitrary function ~\cite{duf}. That is the equation did not fully determine the metric. Perhaps it may be a case similar to the Vaidya solution with an arbitrary function of time where it refers to radial flux of radiation. It was also shown that the fine-tuning conflicted with the staticity of spacetime and thereby with the Birkhoff's theorem ~\cite{duf}. A detailed analysis with thermodynamical properties of black holes in this setting is studied in a number of papers, for instance ~\cite{zan}, and an extension to black string in the braneworld scenario has also been considered ~\cite{kas}. In all the cases, the study is essentially confined to one higher dimension or the bulk-brane system. Ours is slightly different case where we have a Kaluza-Klein split up, ${\ma M}^4 \times {\ma K}^{n-4}$ with the extra dimensional space of constant curvature. We then prescribe the double fine-tuning which in addition relates the radius of curvature of extra dimensional space, ${\ma K}^{n-4}$, to $\Lambda$ and $\alpha$. This makes the four dimensional part of the equations vacuous leaving only one scalar equation following from the extra dimensional part. But it determines the metric fully with two arbitrary constants. It is remarkable that in this case the fine-tuning of the parameters does not conflict with the staticity of spacetime. \\ 

The question is, do there exist other analogues of Vaidya, NUT and Kerr? The analogues of Vaidya and NUT have been obtained and they follow in a straightforward way ~\cite{md2} and so does the analogue of radiating NUT which we have obtained in the following. However it is not possible to obtain the Kerr analogue in this setting because axial symmetry of the Kerr geometry is not compatible with the spherical symmetry of space of constant curvature of ${\ma K}^{n-4}$. We shall also consider the lower dimensional solutions representing mass point with dilaton in two dimension and BTZ black hole and its generalization in three dimension. To get the feel of the physical behavior of the black hole solution ~\cite{md1} and the other solutions, we shall study asymptotic behavior of the $4$-dimensional stresses at both ends ($r\to0, r\to\infty$).\\       

The paper is organized as follows. In the next section, we set up the Kaluza-Klein split up of the field equations which is followed by the discussion of exact solutions. In Secs. 4 and 5, we do the asymptotic expansion of stresses for the black hole spacetime and we end with a discussion. 

\section{Kaluza-Klein for Einstein-Gauss-Bonnet gravity}
Following ~\cite{md1,md2}, we write the action for $n$-dimensional spacetime
\begin{equation}
S=\int d^nx \sqrt{(-g)}\left(\frac1{2\kappa_n^2}(R-2\Lambda+\alpha L_{GB}\right)+ S_{\rm matter}
\end{equation}
where $\kappa_n\equiv\sqrt{8\pi G_n}$, $G_n$ is the $n$-dimensional gravitational constant, $\alpha$ is the Gauss-Bonnet coupling constant, $R$ is the $n$-dimensional Ricci scalar and $\Lambda$ is  the cosmological constant. Note that the quadratic Gauss-Bonnet term makes no contribution to the equation of motion for  $n\leq 4$ and hence this will be the proper action for $n=5,6$. For $n>6$, we should include the cubic term in the Lovelock Lagrangian ~\cite{lovelock}. The Gauss-Bonnet term is given by
\begin{equation}
L_{GB}\equiv R^2-4 R_{\mu\nu}R^{\mu\nu}+R_{\mu\nu\rho\sigma}R^{\mu\nu\rho\sigma}.
\end{equation}

We are going to consider $\alpha\geq 0$ because in some superstring models it is related to the string tension which is taken to be positive.
The gravitational equations following from this action reads as follows: 
\begin{equation}
{\ma G}^\mu_{~~\nu} \equiv {G}^\mu_{~~\nu} +\alpha {H}^\mu_{~~\nu} +\Lambda \delta^\mu_{~~\nu}=\kappa_n^2 {T}^\mu_{~~\nu}, \label{beq}
\end{equation}
where 
\begin{equation}
{G}_{\mu\nu}\equiv R_{\mu\nu}-{\frac12}g_{\mu\nu}R
\end{equation}
is the Einstein tensor and
\begin{eqnarray}
{H}_{\mu\nu}&\equiv&2\Bigl(RR_{\mu\nu}-2R_{\mu\alpha}R^\alpha_{~\nu}-2R^{\alpha\beta}R_{\mu\alpha\nu\beta}\Bigr. \nonumber \\
&&~~~~\Bigl.+R_{\mu}^{~\alpha\beta\gamma}R_{\nu\alpha\beta\gamma}\Bigr)
-{\frac12}g_{\mu\nu}{L}_{GB}. 
\end{eqnarray}
Here $T_{\mu\nu}$ is the energy-momentum tensor following from the matter Lagrangian. For $n\leq 4$, the linear Einstein-Hilbert term suffices because ${H}_{\mu\nu}=0$ then.\\

We consider $n$-dimensional Kaluza-Klein vacuum spacetime with $T_{\mu\nu} = 0$, which is locally homeomorphic to ${\ma M}^{d} \times {\ma K}^{n-d}$ with the metric, $g_{\mu\nu}=\mbox{diag}(g_{AB},r_0^2\gamma_{ab})$, $A,B = 0,\cdots,d-1;~a,b = d, \cdots,n-1$. Here $g_{AB}$ is an arbitrary Lorentz metric on ${\ma M}^d$, $r_0$ is a constant and $\gamma_{ab}$ is the unit metric on the ($n-d$)-dimensional space of constant curvature, ${\ma K}^{n-d}$. \\

In this paper, we shall set $n=6$ and $2\leq d\leq4$. We consider the split up of a six dimensional spacetime into two, three and four dimensional spacetime and a space of constant curvature. The interesting feature of this split up is that it brings in Gauss-Bonnet effects down on dimension $\leq4$. In particular, we also consider the case where ${\ma K}^4$ decomposes as a product of two unit spaces of constant curvature, ${\ma K}^4={\ma P}^2\times{\ma Q}^2$. It turns out that ${\ma G}^\mu_{~~\nu}$ for ${\ma M}^{d} \times {\ma K}^{n-d}$ also gets decomposed as follows ~\cite{md2}:
\begin{eqnarray}
&&{\ma G}^A_{~~B}=\biggl[1+\frac{2{\bar k}\alpha(n-d)(n-d-1)}{r_0^2}\biggl]\overset{(d)}{G}{}^A_{~B}+ \\
&& \hspace*{2em} \alpha \overset{(d)}{H}{}^A_{~B}+\biggl[\Lambda-\frac{{\bar k}(n-d)(n-d-1)}{2r_0^2}\nonumber \\
&&\hspace*{2em} \left(1+ \frac{{\bar k}\alpha(n-d-2)(n-d-3)}{r_0^2}\right)\biggl] \overset{(d)}{\delta}{}^A_{~B},\label{dec1}  \\[2ex]
&&\hspace*{-1.5em} {\ma G}^a_{~~b}=\delta^a_{~~b}\biggl[-\frac12\overset{(d)}{R}+\Lambda-\frac{(n-d-1)(n-d-2){\bar k}}{2r_0^2} \nonumber \\
&&\hspace*{.7em} -\alpha\biggl\{\frac{{\bar k}(n-d-1)(n-d-2)}{r_0^2}\overset{(d)}{R}+\frac12 \overset{(d)}{L}_{GB}\biggl\}\biggl].\label{dec2}
\end{eqnarray}  

The vacuum equation ${\ma G}^A_{~~B}=0$ is a tensorial equation on ${\ma M}^{d} $, while ${\ma G}^a_{~~b}=0$ is a scalar constraint. The former is the usual Einstein equation with $\Lambda$ redefined while the latter then works as a constraint on it. For $d\leq 4$, note that $\overset{(d)}{H}{}^A_{~B}=0$. We shall consider the cases $d=2,3$ separately and shall now take up the case $d=4$. The former equation (\ref{dec1}) is simply the Einstein equation with $\Lambda$ redefined and hence it would have Schwarzshild-dS/AdS as the general static solution. When this is put into the scalar equation (\ref{dec2}), it turns into dS/AdS. It prohibits existence of mass point on  ${\ma M}^{4}$. However there is an interesting special case in which it is made vacuous by choosing the parameters such that both square brackets in equation (\ref{dec1}) vanish. For $d=4, n=6$, this means $\bar k/r_0^2=-1/4\alpha=\Lambda$. Then we are left only with the scalar equation (\ref{dec2}). All the solutions (but for one three dimensional case) we shall be considering in the next section will refer only to this one equation. Note that with this choice of parameters, ${\ma M}^{4}$ can harbor no matter at all as ${\ma G}^A_{~~B} = 0$. This prescription also implies that for $\alpha > 0$, ${\bar k} = -1$ and $\Lambda < 0$ and for $\alpha < 0$, ${\bar k} =1$ and $\Lambda > 0$. Hereafter we  obtain the vacuum solution with $T_{\mu\nu} = 0$ for this setting. The governing equation is then a single scalar equation on ${\ma M}^4$, $\ma G^{a}_{~~b} = 0 $, which is equivalent to $T=0$ given by 
\begin{equation}
 \overset{(d)}{R}+ \frac{\alpha(n-d)}{2}\overset{(d)}{L}_{GB}+\frac{2(n-d)-3}{\alpha(n-d)(n-d-1)} =0\,.\label{scalarequation}
\end{equation} 
Then, solutions on ${\ma M}^{d} \times {\ma K}^{n-d}$  are obtained by solving  this equation alone for $g_{AB}$. Since this is the trace zero, $T=0$, condition and hence it will also remain satisfied for a trace free distribution. In particular, if we transform a static solution to retarded/advanced time coordinate and make the parameters of the solution function of the transformed time, the above equation will remain undisturbed. That is the only governing equation and we shall therefore have radiating analogue of static solution. That is, all static solutions could be made to Vaidya radiate. This is a remarkable general feature of this construction.  

\section{Exact solutions}
\subsection{$d=2$, $n=6$}
For $d=2$ and ${\ma M}^{2} \times {\ma K}^{4}$, from \cite{md2} we have,
\begin{eqnarray}
&& r_0^2=\frac3\Lambda\bar k\left(1\pm\sqrt{1+\frac43\alpha\Lambda}\right),\nonumber \\
&& \overset{(2)}{R}=\frac{2r_0^2}{r_0^2+12\alpha\bar k}\left(\Lambda-\frac{3\bar k}{r_0^2}\right), \label{constM2}
\end{eqnarray}
where $\bar k/r_0^2$ is the constant curvature of ${\ma K}^{4}$ with $\bar k = \pm 1$. It then further solves to give
$$\overset{(2)}{R}=C=2\Lambda\frac{\sqrt{9+12\alpha\Lambda}}{\sqrt{9+12\alpha\Lambda}\pm3(1+4\alpha\Lambda)}.$$ 
On the other hand for the case ${\ma M}^{2} \times {\ma K}^2\times {\ma Q}^2$ where ${\ma K}^2$ and ${\ma Q}^2$ are spaces of constant curvature $R_1$ and $R_2$ respectively, the equations (\ref{dec1}) and (\ref{dec2}) take the form,
\begin{eqnarray}
& & \hspace*{-2em} 0 = \left(\Lambda  - \alpha R_1R_2 - {\displaystyle \frac{R_1+R_2}{2}} \right)\,\delta^A_B\\
& &\hspace*{-2em} 0 = \left(\Lambda  - \alpha\overset{(2)}{R}R_2 - {\displaystyle \frac{\overset{(2)}{R}+R_2}{2}} \right)\,\delta^a_b\\
& & \hspace*{-2em} 0 = \left(\Lambda  -\alpha\overset{(2)}{R}R_1 - {\displaystyle \frac{\overset{(2)}{R}+R_1}{2}} \right)\,\delta^c_e
\end{eqnarray}
where $A,B=1,2$, $a,b=3,4$ and $c,e=5,6$ and $\overset{(2)}{R}$ is the curvature of ${\ma M}^{2}$. These equations determine $\Lambda$ and $\overset{(2)}{R}$ as
\begin{equation}
\Lambda=-\frac{1}{4\alpha}, \qquad\overset{(2)}{R}=-\frac{1}{2\alpha}\label{curvd2}
\end{equation}
and we also have 
\begin{equation}
0=1+2\alpha(R_1+R_2)+4\alpha^2R_1R_2. 
\end{equation}
For $R_1=-R_2$ we obtain $R_1=\pm 1/2\alpha$ while for $R_1=R_2$, $R_1=-1/2\alpha$ and  $\overset{(2)}{R}= C = -1/2\alpha$ always. \\ 

However the solution on ${\ma M}^{2}$ in the two cases turns out to be the same and it is given by 
\begin{equation}
ds^2=\frac{1}{f(r)} dr^2-f(r) dt^2
\end{equation}
is
\begin{equation}
f(r) = -\frac12 C r^2+C_1r+C_2
\end{equation}
where $C_1$ and $C_2$ are the constants of integration. This is the two dimensional dilatonic mass point solution ~\cite{ms}. Here $C$ refers to the cosmological constant while $C_1$ to the mass and $C_2$ is determined in terms of the other two by the dilaton dynamics ~\cite{mm}. In this case, transformation to retarded/advanced time is rather trivial as the curvature does not involve time derivative. 

\subsection{$d=3$, $n=6$}
In this case we write for the equations (\ref{dec1}) and (\ref{dec2}),
\begin{eqnarray}
&&{\ma G}^A_{~~B}=\biggl[1+\frac{12{\bar k}\alpha}{r_0^2}\biggl]\overset{(3)}{G}{}^A_{~B}+ 
\biggl[\Lambda-\frac{3{\bar k}}{r_0^2} \biggl] \overset{(3)}{\delta}{}^A_{~B},\label{deca1} \\[2ex]
&&{\ma G}^a_{~~b}=\delta^a_{~~b}\biggl[-\frac12\overset{(3)}{R}+\Lambda-\frac{\bar k}{r_0^2} 
-\alpha\frac{2\bar k}{r_0^2}\overset{(3)}{R}\biggl].\label{deca2}
\end{eqnarray}
As before setting the coefficients in (\ref{deca1}) to zero we obtain $\bar k/r_0^2=-1/12\alpha=\Lambda/3$. This implies if $\alpha>0$, $\bar k<0$ and $\Lambda <0$, and the other equation then gives $2\Lambda= \overset{(3)}{R}$. The vacuum solution is then given by 
$$ds^2=\frac{1}{f(r)} dr^2+r^2 d\phi^2-f(r) dt^2$$
where
\begin{equation} 
f(r)=-\frac{\Lambda}{3}r^2+\frac{q}{r}-M 
\end{equation}
where $q$ and $M$ are integration constants. The Einstein tensor for this reads as 
\begin{equation}
G^1_1=G^3_3=-(\frac\Lambda3+\frac{q}{2r^3}),\qquad G^2_2=-\frac\Lambda3+\frac{q}{r^3}.
\end{equation}
If we take $\Lambda<0$, this is the three dimensional BTZ  black hole ~\cite{btz} of mass $M$ sitting in a traceless distribution generated by the parameter $q$. This is a new generalized BTZ solution similar to the one obtained in Ref. ~\cite{bdk} where BTZ black hole was sitting in a string dust distribution (There also exists a BTZ-like-stringy solution on $2$-brane in $5$ dimensional Gauss-Bonnet gravity ~\cite{left}). The new charge $q$ is caused by Kaluza-Klein split up. It is an interesting new black hole solution. \\

It can also be made to Vaidya radiate by transforming the metric to retarded/advanced time, $v$ and writing $M=M(v), q=q(v)$. The metric will then read as 
\begin{equation}
ds^2=2dvdr + r^2 d\phi^2-f(v,r) dt^2 
\end{equation}
where $f$ is the same as above with $M(v), q(v)$. This is the Vaidya generalization and the stresses would now read as
\begin {eqnarray}
&&G^1_1=G^3_3=-(\frac\Lambda3+\frac{q}{2r^3}),\nonumber\\ && \hspace*{-1em} G^2_2=-\frac\Lambda3+\frac{q}{r^3}, ~~ G^1_2=\frac{\dot{M}r-\dot{q}}{2r^2}.
\end{eqnarray}
This happens because the equation (\ref{deca2}) continues to remain satisfied with $M, q$ being functions of the time $v$. \\ 

Without the above prescription of the parameters, the equation (\ref{deca1}) admits the well-known BTZ black hole solution,
$$ f(r)=-\lambda r^2-M$$ 
with redefined $\lambda$ given by
$$\lambda=\frac{\Lambda-\bar k/r_0^2}{1+12\alpha\bar k/r0^2}.$$
Then the equation (\ref{deca2}) determines $\Lambda=4\frac{\bar k}{r_0^2}\left(1-3\alpha\bar k/r_0^2\right)$ and consequently $\lambda=3(1-4\alpha\bar k/r_0^2)/(1+12\alpha\bar k/r_0^2)$. It is no surprise that we have BTZ solution because we solve the same equation. It is worth noting that the scalar constraint (\ref{deca2}) in this case simply determines $\lambda$ in terms of $\alpha$ and the constant curvature of ${\ma K}^{3}$ and does not disturb the solution otherwise as it does in $4$-dimensional case where Schwarzschild mass is knocked off. That is why we have interesting solutions in both cases with the prescription, $\bar k/r_0^2=-1/12\alpha=\Lambda/3$ and without.\\ 

However unlike the earlier solution with $q$, this solution cannot Vaidya radiate because it has in addition also to satisfy the equation (\ref{deca1}) which is not possible. It is interesting that though BTZ black hole cannot radiate but the generalized BTZ with additional gravitational charge $q$ can. This is  because it has only to satisfy the scalar constraint alone coming from the extra dimensional equation~(\ref{deca2}). \\ 
  
From now onwards, we shall take $d=4, n=6$ and $\bar k/r_0^2=-1/4\alpha=\Lambda$. In the next subsections, we shall first recall the analogues of Schwarzschild, Vaidya and NUT solutions obtained in Ref. ~\cite{md2} and shall then make NUT solution radiate as well as show that there exists no Kerr analogue in this setting.

\subsection{Static Schwarzschild-like solution}
We seek a spherically symmetric static solution of the equation (\ref{scalarequation}) for the metric 
\begin{eqnarray}
ds^2&&=\frac{1}{f(r)} dr^2+r^2 d\Omega^2-f(r) dt^2
\end{eqnarray}
which is given by as in Ref. ~\cite{md1},
\begin{equation}
f(r)= 1 + \frac{r^{2}}{4\alpha}\left[ 1\pm  \sqrt{\frac23 + 16\left( \frac{\alpha^{3/2}  M}{r^{3}}
  - \frac{\alpha ^{2}  q}{r^{4}}\right)}\right] \label{Scharzschildsol}
\end{equation}
where $M$ and $q$ are arbitrary dimensionless constants and they are normalized by $\alpha$. The former is mass of the black hole while the latter is new Kaluza-Klein gravitational charge ~\cite{dmpr}. 

\subsection{Vaidya-like solution} 
It turns out that if we transform the previous solution to retarded/advanced time coordinate and make constants $M$ and $q$ functions of the time coordinate $v$, it continues to be a solution of the equation (\ref{scalarequation}). The Vaidya analogue is therefore given by,
\begin{eqnarray}
ds^2=&& -f(v,r)dv^2 +2dvdr+ r^{2}d\Omega^{2}\nonumber \\
&& =(\omega^1)^2+(\omega^2)^2+(\omega^3)^2-(\omega^4)^2,
\end{eqnarray}
where $$\omega^1=\frac{dr}{\sqrt{f(v,r)}},\, \omega^2=r\,d\theta,\,\omega^3=r\sin\theta d\varphi,$$ $$\omega^4=\sqrt{f(v,r)}\,dv-\frac{dr}{\sqrt{f(v,r)}}$$
and $f$ is as given previously in (\ref{Scharzschildsol}) with $M=M(v)$ and $q=q(v)$.

\subsection{NUT-like solution}
Another generalization of the solution (\ref{Scharzschildsol}) is a NUT-like solution. For the metric
\begin{eqnarray}
ds^2= && \frac{dr^2}{
f(r)}  + (r^{2} + l^{2})d\Omega^{2}-f(r)(2  l  \cos\theta d\varphi + 
dt)^{2}\nonumber \\
&& =(\omega^1)^2+(\omega^2)^2+(\omega^3)^2-(\omega^4)^2,
\end{eqnarray}
where $$\omega^1=\frac{dr}{\sqrt{f(r)}},\, \omega^2=\sqrt{r^2+l^2}\,d\theta, \omega^3=\sqrt{r^2+l^2}\sin\theta d\varphi,$$ $$\omega^4=\sqrt{f(r)}(2l\cos(\theta)\,d\varphi+dt).$$ 
Then equation (\ref{scalarequation}) solves to give 
\begin{eqnarray}
& &\hspace{-1.5em} f(r)=\frac{
(r^{2} + l^{2})(r^{2}+ l^{2} +4\alpha)}{4\alpha(r^{2} - 3  l^{2})}\pm  \label{Taub-nut} \\[1ex] && \frac{ \sqrt{(r^2 + l^{2})\left[4\alpha ^{3/2}(r^{2} - 3l^{2})(Mr - \alpha^{1/2}q) + A(r)\right]}}{2\alpha(r^{2} - 3  l^{2})}\nonumber
\end{eqnarray}
and
\begin{eqnarray}
&& \hspace{-3em} A(r)= \frac{r^4}{6} (r^2+3l^2) +  \\[1ex] &&\hspace*{1em} l^2 \left(\frac{l^2}{4}(9r^2+l^2)+
 2\alpha (5r^2+l^2)+16\alpha^2\right).\nonumber
\end{eqnarray}
It reduces to (\ref{Scharzschildsol}) for $l=0$.

\subsection{Radiating NUT-like solution}
As before, we can make NUT solution radiate by the same method of transforming it to the retarded/advanced time coordinate, and interestingly again it continues to be solution of the scalar constraint equation (\ref{scalarequation}). We write the metric 
\begin{eqnarray}
&& \hspace{-1em} ds^2= -f(v,r)(dv + 2  l\cos\theta d\varphi)^2 + 2dr(dv + 2  l\cos\theta d\varphi) +\nonumber \\ &&  (r^{2} + l^{2})d\Omega^{2}=(\omega^1)^2+(\omega^2)^2+(\omega^3)^2-(\omega^4)^2, 
\end{eqnarray}
where $$\omega^1=\frac{dr}{\sqrt{f(v,r)}},\, \omega^2=\sqrt{r^2+l^2}\,d\theta,\, \omega^3=\sqrt{r^2+l^2}\sin\theta d\varphi,$$ $$\omega^4=\sqrt{f(v,r)}(dv+2l\cos\theta\,d\varphi)-\frac{dr}{\sqrt{f(v,r)}}.$$ 
Now $f(v,r)$ is the same as given before for NUT solution (\ref{Taub-nut}) with $M$ and $q$ being functions of $v$. It reduces to Vaidya solution when $l=0$. This is a new solution which was not obtained in Ref. \cite{md2}. 

\subsection{Non existence of a Kerr analogue solution}
We seek a stationary solution for a rotating mass point, an analogue of Kerr spacetime, with the metric on ${\ma M}^4$ reading as:
\begin{equation}
ds^2=-\frac{f(r)^2}{\rho^2}\left(dt-a\sin^2\theta d\varphi\right)^2+\frac{\rho^2}{f(r)^2}dr^2+$$
$$\rho^2d\theta^2+\frac{\sin^2\theta}{\rho^2}\left((r^2+a^2)d\varphi-adt\right)^2, \label{metric} 
\end{equation}
where $\rho^2=r^2+a^2\cos^2\theta$ and $f(r)$ is an arbitrary function of $r$. Here $a$ is the parameter defining axial symmetry.

As usual, we have to solve the equation (\ref{scalarequation}) and for which we first compute 
\begin{equation}
\overset{(4)}{R}=   \frac{2-(f^2)''}{\rho ^{2}}
\end {equation}
and 
\begin{eqnarray}
&&\overset{(4)}{L}_{GB} =  \frac{4 F''(r^{2} - 3a^{2}\cos^2\theta)}{
\rho ^{8}} -  \\
&&  \frac{32F'(r^2 - 5a^2\cos^2\theta)}{\rho ^{10}}+\nonumber  \\
&& \frac{16F(5a^4\cos^4\theta- 38r^2a^2\cos^2\theta + 5r^4)}{\rho ^{12}}  -\nonumber  \\ &&\frac{8a^2(3a^4\cos^6\theta(a^2+3r^2)-5a^2\cos^4\theta(9a^2r^2-6r^4-}{}\nonumber  \\ &&\frac{a^4)+r^2\cos^2\theta(38a^4+ 45a^2r^2+9r^4)- 3r^6-5r^4a^2}{\rho ^{12} }\nonumber
\end{eqnarray}  
where $F(r)=f^2/2-(r^{2} + a^{2})f.$
With this angular dependence of $\overset{(4)}{L}_{GB}$, it is clear that there exists no solution for the equation (\ref{scalarequation}). This is perhaps because of incompatibility of the axial symmetry of ${\ma M}^4$ with the constant curvature of ${\ma K}^2$. Thus there can exist no Kerr analogue in this setting. 

\section{Asymptotic behavior of metric and stresses}
We are  now interested in studying asymptotic behavior of the metric and the stresses implied by the four dimensional solutions. This is to get the physical feel of Gauss-Bonnet effects descending down to four dimension. We shall consider the limit at both ends, $r\to\infty$ and $r\to0$. We begin with the former asymptotic limit.  
\subsection{Schwarzschild-like solution}
The metric function $f(r)$ asymptotically goes as:
\begin{equation}
 f_{\rm as}(r)\approx 1 + \frac{r^{2}}{4\alpha}\left(1\pm  \sqrt{\frac23}\right)\pm \sqrt{6}\left(\frac{\alpha^{(1/2)}Mr - \alpha   q}{r^{2}}\right).
\label{Schbh}
\end{equation}
 The stresses go as 
\begin{equation}
 G_{11} =-G_{44}\approx \lambda+  \frac{\beta_0}{r^{4}},\quad G_{22} =G_{33}\approx \lambda  -  \frac{\beta_0}{r^{4}}
\end{equation}
and $\lambda$ and $\beta_0$ are 
\begin{equation}
  \lambda  =\frac{3}{4\alpha}\left(1\pm \sqrt{\frac {2}{3}
}\right)  \label{lambda}
\end{equation}
\begin{equation}
\beta_0 =\pm \alpha   q  \sqrt{6}
\end{equation}
where $\pm$ corresponds respectively to the two solutions with $\pm$ sign in the metric function $f(r)$. It clearly indicates a behavior similar to RN-AdS. This is a black hole of mass $M$ with additional Kaluza-Klein charge $q<0$ ~\cite{dmpr} sitting in an anti-deSitter spacetime. 

\subsection{Vaidya-like solution}
The metric function $f(v,r)$ has the same behavior as in (\ref{Schbh}) with now $M(v)$ and $q(v)$, and the stresses read as 
\begin{eqnarray}
&& G_{11}\approx \lambda +\frac{\beta_0}{r^4}+G_{14},\quad G_{44}\approx -\lambda -\frac{\beta_0}{r^4}+G_{14},\nonumber\\ &&  G_{22}=G_{33}\approx   \lambda  - \frac {\beta_0}{r^{4}} 
\end{eqnarray}
and 
\begin{equation}\hspace*{-1.5em}
G_{14}=  \mp\frac{3\sqrt{6\alpha}\dot{M}(v)}{ \lambda}\frac1{r^4}
\end{equation}
where $\lambda$ and $\beta_0$ being the same as before. The radiation density, given by $G_{14}$ gets added to Schwarzschild stresses $G_{11}$ and $G_{44}$.  
\subsection{NUT-like solution}
The metric function (\ref{Taub-nut}) asymptotically reads as:
\begin{eqnarray}
&&\hspace{-2em} f_{\rm as}(r)\approx 1+\frac{r^{2}+5l^2}{4\alpha}\left(1\pm
 \sqrt{\frac{2}{3}}\right)
  \pm \nonumber \\  &&\hspace*{1em} \sqrt{6}\left(\frac{\alpha^{1/2}  M(r^2-l^2)}{r^3}-   
  \frac{\alpha   q}{r^{2}} \right)\pm \nonumber \\  && \hspace*{1em}
\frac{l^2}{r^2}\left(5\sqrt{\frac{3}{2}}\left(\frac{17}{24}\frac{l^2}{\alpha}  + 1\right) + 4  \left(\frac{l^{2}}{\alpha} + 
1\right)\right)\!.  \label{NUT-as}
\end{eqnarray}
It reduces to Schwarzschild limit for $l\rightarrow 0$ and the stresses have the same Schwarzschild behavior with unchanged $\lambda$ but 
 $\beta$ is now 
\begin{equation}
 \beta= \beta_0 - l^{2}\left(6\pm 5 \sqrt{ \frac {3}{2}}  \right)- \frac {l^{4}}{\alpha} \left(6\pm \frac{39}{8}\sqrt{\frac32}    \right).  \label{beta} 
\end{equation}
We recover $\beta_0$ for $l\rightarrow 0$.

\subsection{Radiating NUT-like solution}
The expression for the metric function $f(v,r)$ is the same as in (\ref{NUT-as}) with $M(v)$ and $q(v)$ and the stresses read as 
\begin{equation}
G_{11}\approx  \lambda +\frac{\beta}{r^4}+G_{14},\quad G_{44}\approx  -\lambda -\frac{\beta}{r^4}+G_{14}, 
\end{equation}
\begin{equation}
G_{14}\approx \mp 3 \frac{\sqrt{6\alpha}}{\lambda}\frac{\dot{M}(v)}{r^4},   
\end{equation}
\begin{equation}
G_{22}=G_{33}\approx  \lambda -\frac{\beta}{r^4},\quad G_{12}=G_{24}\approx 0
\end{equation}
\begin{equation}
G_{13}\approx \mp 3\,l \sqrt{\frac{2\alpha}{\lambda}}\frac{\cos\theta}{\sin\theta}\frac{\dot{M}(v)}{r^4}.
\end{equation}
Both $\lambda$ and $\beta$ are the same as for NUT.

\section{Behavior of metric and stresses for  $r\rightarrow0$}
We take the limit on the other end and write the two lowest order terms. 
\subsection{Schwarzschild} 
\begin{equation}
f(r)\approx 1 \pm \sqrt{\frac{Mr}{\sqrt{\alpha}}-q}
\end{equation}
\begin{equation}
G_{11}=-G_{44} \approx \pm \frac {1}{r^2}\sqrt{\frac{2Mr}{\sqrt{\alpha }}-q}, 
\end{equation}
\begin{equation} 
G_{22}=G_{33}\approx\pm\frac18\frac{M}{(\alpha)^{1/4}}\frac1r\frac{3Mr-4q\sqrt{\alpha}}{(Mr-q\sqrt{\alpha })^{3/2}}.
\end{equation}
\subsection{Vaidya} 
It is the same as Schwarzschild except for $M=M(v), q=q(v)$, and $G_{11}\to G_{11}+G_{14}$ and $G_{44}\to G_{44}+G_{14}$ where $G_{14}$ is given by 
$$G_{14}\approx\mp\frac1r\frac1{1\pm \sqrt{rM(v)/\sqrt{\alpha}-q(v)}}\partial_v\sqrt{\frac{M(v)}{\sqrt{\alpha }}r-q(v)}.$$
\subsection{NUT}
\begin{equation}
f(r)\approx  - \frac {l^{2}}{12\alpha} -  \frac {1}{3}  \pm \frac {
\sqrt{X}}{12\,\alpha}  \mp \frac {2M
\sqrt{\alpha }r}{\sqrt{X}}, 
\end{equation}
where $X$
\begin{equation}
X = 16\,\alpha ^{2}(3q + 4) + l^{2}(8\alpha+l^{2} ).
\end{equation}
and 
\begin{equation}
G_{44}\approx k_1+  k_2 \frac{r^2}{l^2}\quad G_{22} \approx 0,
\end{equation}
where 
\begin{eqnarray}
&&\hspace*{-2em} k_1=\frac{ 8\alpha - l^{2}  \pm X^{1/2}}{12l^{2}\alpha }\quad\mbox{and}\quad
k_2=\frac {4(\alpha +  l^2) }{9l^2\alpha}\mp  \nonumber \\  && \frac{25l^2}{36\alpha X^{1/2}}\mp\frac{4\alpha(33q + 46)}{9l^{2}X^{1/2}} \mp  \frac{ 38}{9 X^{1/2}}\pm  \frac{24\alpha^2 M^2}{X^{3/2}}.\nonumber\end{eqnarray}
It is worth noting that in the limit $l\rightarrow 0$, $f(r)$ in NUT solution does not go over to the corresponding Schwarzschild limit. This is because there is a continuity problem for the function $f(l,r)$ at the point $(l,r)=(0,0)$. That is, $\lim_{l\rightarrow0}\lim_{r\rightarrow0}f(l,r)\neq\lim_{r\rightarrow0}\lim_{l\rightarrow0}f(l,r)$. \\ 

The remarkable feature of the limit $r\to0$ is that the metric always remains regular and stresses have softer divergence in all the cases.

\section{Discussion} 
The gravitational equation also splits up in accordance with the spacetime split up into $d$ and $(n-d)$ dimensional part. The remarkable feature of the split up is to bring Gauss-Bonnet effect down to ${\ma M}^d$ even for $d\leq4$. In general, the equation (\ref{dec1}) on ${\ma M}^d$ is the usual Einstein equation with a redefined $\Lambda$ which would in general solve to give Schwarzschild-dS/AdS. This is when put into the constraint equation (\ref{dec2}) which then demands $M=0$, leaving the solution to be dS/AdS. However there is an interesting exceptional case in which the equation (\ref{dec1}) becomes vacuous and we are only left with the single equation (\ref{scalarequation}) which then entirely governs the dynamics of ${\ma M}^d$. All the $4$ dimensional solutions refer to this setting which is characterized by the prescription, $\bar k/r_0^2=-1/4\alpha=\Lambda$. \\ 

The most interesting solution ~\cite{md1} is however the static black hole with mass and a Maxwell-like gravitational charge which asymptotically resembles RN-AdS. Here we have a Maxwell-like charge without Maxwell field because the solution is obtained by solving the equation (\ref{scalarequation}) which is the trace zero condition. The Maxwell-like charge is a characteristic of tracelessness in $4$ dimensional static spacetime. It has been envisioned as a black hole being created out of pure curvature ~\cite{dm,d}. As the prescription of the parameters prohibit existence of any matter field on ${\ma M}^d$, all the parameters in the solution therefore arise from the extra dimensional scalar equation signifying $T=0$. Further all the solutions could be made to Vaidya radiate and there is all pervading AdS background. Let us imagine collapse of Vaidya null dust in AdS spacetime giving rise to a static black hole when collapse ends. The interior of the static solution is the collapsing null dust. A black hole could thus be thought of being formed out of an AdS spacetime ~\cite{dm,d}. It is remarkable that all the solutions including the lower dimensional generalized BTZ black hole share this interesting feature. It is interesting that analogues of all the static black hole solutions (Schwarzschild, NUT and generalized BTZ) could in fact be formed by collapse of Vaidya null dust in AdS. This is a very important general feature of this setting. These are all examples of trace free matter being created by pure curvature. This is in some sense manifestation of ``matter without matter'' ~\cite{dm,d}.\\

% {That is, here is a prescription all these objects could thus be thought of All  It is remarkable that all the static solutions including the generalized can be made to Vaidya radiate and there exists an AdS background.  This is a general feature of the setting characterized by the property that $d$-dimensional equation is completely vacuous and it is entirely determined by the scalar constraint coming from the extra dimensional equation. Thus all the solutions including the two dimensional mass point with dilaton and the generalized BTZ black hole in tracefree distribution (the sole exception being BTZ solution) are similarly pure curvature creations, examples of ``matter without matter''. It was shown that the four dimensional static black hole could be envisioned to have formed by the collapse Vaidya null dust from an AdS spacetime ~\cite{dm,d}. The new generalized BTZ black hole found here could similarly be thought of as formed from an AdS. We thus have new examples of pure curvature creation. It turns out that the governing equation can always be solved for spherical symmetry and but it has no Kerr analogue solution for axial symmetry. This is because of incompatibility of Kerr symmetry with that of space of constant curvature of ${\ma K}^{2}$. 
%Besides two and three dimensional solutions, the radiating NUT-Vaidya solution is also new. %It is worth noting that Schwarzschild and NUT solutions when transformed to retarded/advanced time coordinate, they continue to remain solutions of the scalar equation, ((\ref{scalarequation}).}  

One of the aims of this investigation was to study the behavior of the black hole metric and its stresses at the two ends for large and small $r$. For the large end, it tends to behave like an RN-AdS-like spacetime. Note that $q$ is negative and it is a Kaluza-Klein gravitational charge ~\cite{dmpr}. At the other end, it is interesting to see that the metric is regular at $r=0$ and the stresses have softer divergence than otherwise. It is known that Gauss-Bonnet contribution tends to weaken the singularity ~\cite{d1,noo} in the corresponding solution for $n=5,6$. It is remarkable that this desirable feature has been brought down to four dimension. In general we see through all these solutions that Kaluza-Klein contributes an additional gravitational charge while Gauss-Bonnet makes the metric regular at the center and the singularity weaker. It is also interesting that all the static solutions could be made to Vaidya radiate. 

\section*{Acknowledgement} 
We wish to thank the anonymous referee for the useful comments. AM wishes to thank IUCAA for warm hospitality. The work of A.M.\ is supported by Ministerio de Ciencia y
Tecnolog\'{\i}a,  FIS2007-63034, and Generalitat de Catalunya, 2005SGR-00515 (DURSI).

%======================================%
%<<<<<<<<<<<<< REFERENCES >>>>>>>>>>>>>%
%======================================%

\end{document}